# Cooperative Multi-Agent Reinforcement Learning Based Distributed Dynamic Spectrum Access in Cognitive Radio Networks

Xiang Tan, Li Zhou, Haijun Wang, Yuli Sun,
Haitao Zhao, *Senior Member, IEEE,* Boon-Chong Seet, *Senior Member, IEEE,*
Jibo Wei, *Member, IEEE,* and Victor C. M. Leung, *Life Fellow, IEEE*

**Abstract**

With the development of the 5G and Internet of Things, amounts of wireless devices need to share the limited spectrum resources. Dynamic spectrum access (DSA) is a promising paradigm to remedy the problem of inefficient spectrum utilization brought upon by the historical command-and-control approach to spectrum allocation. In this paper, we investigate the distributed DSA problem for multi-user in a typical multi-channel cognitive radio network. The problem is formulated as a decentralized partially observable Markov decision process (Dec-POMDP), and we proposed a centralized off-line training and distributed on-line execution framework based on cooperative multi-agent reinforcement learning (MARL). We employ the deep recurrent Q-network (DRQN) to address the partial observability of the state for each cognitive user. The ultimate goal is to learn a cooperative strategy which maximizes the sum throughput of cognitive radio network in distributed fashion without coordination information exchange between cognitive users. Finally, we validate the proposed algorithm in various settings through

This work was supported in part by the National Natural Science Foundation of China under Grant 6193000305.
(*corresponding author: Li Zhou.*)

X. Tan, L. Zhou, Y. Sun, H. Wang, H. Zhao and J. Wei are all with College of Electronic Science and Technology, National University of Defense Technology, Changsha, 410073, China (E-mail: {tanxiang, zhouli2035, haijunwang14, sunyuli19, haitaozhao, wjbhw}@nudt.edu.cn).

Boon-Chong Seet is with the Department of Electrical and Electronic Engineering, Auckland University of Technology, Auckland 1142, New Zealand (E-mail: boon-chong.seet@aut.ac.nz).

Victor C. M. Leung is with Shenzhen University, Shenzhen, China and the University of British Columbia, Vancouver, Canada (E-mail: vleung@ieee.org).

extensive experiments. From the simulation results, we can observe that the proposed algorithm can converge fast and achieve almost the optimal performance.

# I. Introduction

The future network is involving into the Internet of Everything. Wireless devices, such as smartphones, wearable fitness recorders and smart home devices are vying for access to the radio spectrum and the number of them is surging. According to Cisco's forecast [1], the number of devices connected to IP networks will reach 29.3 billion by 2023. And by 2030, the demand for wireless access could be 250 times larger than it is today. However, the spectrum resource has almost been fully allocated, and the spectrum management mechanism is lack of efficiency to some extent. It allocates exclusive licensed bands to isolated wireless systems, which guarantees interference-free communication, but it is human-operated. It is difficult for the limited available spectrum to accommodate the growing demand of wireless applications. However, in a given period, some allocated bands are underutilized, while other bands are overwhelmed, which squanders enormous spectrum capacity, and exacerbates the scarcity of spectrum. Therefore, developing a flexible and high-efficiency wireless paradigm of spectrum access to meet the increasing demand is a crucial issue for future wireless communication. Dynamic spectrum access (DSA) technology access the available spectrum flexibly according to the demand of users, which is envisioned as a promising technology to solve the current spectrum inefficiency problems. With the potential advantages of machine learning, we try to propose an intelligent DSA technology.

## *A. Related works*

Some literatures have surveyed the traditional DSA models [2], [3]. However, the traditional model-based approaches focus on designing spectrum access protocols for specific network scenarios, which require prior knowledge of the network dynamics to formulate the wireless networks model and cannot effectively be applied to handle complex real-world problems. The myopic policy [4] and the Whittle index policy [5] are two classic algorithms for single-user scenario when the channels are independent and the state transition matrixes are known in prior. The myopic policy always selects the channel which maximizes the immediate reward and ignores its long-term effects. Only when the state transitions are positively correlated or slightly negatively correlated, the performance is optimal, otherwise is not. When all the channels are independent but not identical, the Whittle index policy can achieve optimal result. When channels follow identical distribution, the Whittle index policy falls into the same round-robin structure as the myopic policy. But they are all not competent for complicated communication scenarios with unknown network dynamics. The development of artificial intelligence (AI) facilitates intelligent wireless communications [6], and machine learning is a powerful tool which is envisioned as a potential solution of DSA problem [7]. Deep reinforcement learning (DRL) [8] embraces the multi-dimensional perception ability of deep neural network (DNN) and the autonomous decision-making ability of reinforcement learning [9]. The agents learn by trial-and-error interaction with its surrounding environment. DRL can be model-free and does not require the network dynamics in prior. At each time step, the agent observes the state of the environment and takes an action according to its own policy, meanwhile receives a scalar reward that evaluates the quality of the action, and then adjusts its strategy to achieve the optimal policy. Due to the curse of dimensionality caused by the exponential growth of the state-action space in the number of states and action variables, the existing works mainly focus on DSA of single-user based on single-agent DRL algorithm. The authors of [10] surveyed the applications of DRL in wireless communications and networking. The work [11] is the pioneer work that implements DRL in the field of DSA, in which the authors formulated the dynamic multi-channel access problem of a single-user as a partially observable Markov decision process (POMDP) and applied Deep Q-Network (DQN) to obtain the optimal access policy in wireless sensor networks, which demonstrated that DQN can outperform the traditional DSA approaches. A deep actor-critic reinforcement learning based framework for dynamic multi-channel access is proposed in [12]. The actor-critic reinforcement learning algorithm consists of a critic network and an actor network. The critic network estimates the Q-value and the actor network updates the policy guided by the Q-value. By decoupling the Q-value estimation and action selection, the actor-critic reinforcement learning based DSA algorithm converges faster and can achieve the near-optimal performance compared with the DQN-based algorithm. A DRL-based MAC protocol for heterogeneous wireless networks was proposed in [13], which investigated the scenario of several heterogeneous networks operating different MAC protocols trying to share the time slots of the common channels. It can maximize the sum throughput and achieve proportional fairness among all networks. The authors of [14] investigated the distributed spectrum access in heterogeneous environments with partial observations, where the intelligent users employ deep recurrent Q-network (DRQN) to make access decision based on previous observations. The simulation results demonstrate the effectiveness and robustness of the DRQN-based approach in cognitive radio networks with multiple heterogeneous primary users (PUs). Furthermore, the real-world wireless communication networks can be modeled as multi-agent systems, in which the agents share the common environment and the interaction between agents can be either cooperative, competitive, or mixed [15]. The current works simply apply the single-agent reinforcement learning to the multi-agent context, which may incur non-stationary and non-convergent results. In [16], the authors devised a deep Q-learning spectrum access (DQSA) algorithm based on Dueling-DQN and long short-term memory (LSTM) for multi-user under partial observability, which learns policies from their own ACK signals and each user employs an independent single-agent Q-Learning algorithm. A central trainer trains the neural network for each user respectively, and updates the network parameters periodically. As a result, the overhead for exchanging information can be eliminated. This DQSA algorithm is only applicable to wireless networks with relatively small number of channels and users. In [17], the resource allocation problem of LTE-U small base stations (SBSs) is modeled as a non-cooperative game, in which SBSs aim at predicting a sequence of channel selection, carrier aggregation and spectrum access to achieve weighted fairness with existing Wi-Fi networks and other LTE-U operators over a given horizon. The simulation results showed that the proposed RL-LSTM algorithm can converge to a mixed-strategy Nash equilibrium (NE). The algorithm is model-based and in fact it is a single-agent Q-Learning framework implemented independently on SBSs. A combination of reservoir computing (a special type of recurrent neural network) and DQN is utilized to design spectrum access strategies for secondary users (SUs) in distributive DSA networks in [18]. The SUs make spectrum access decisions relying on their current and past spectrum sensing outcomes. The algorithm also adopts an independent-learner manner, which does not take the other users' policy into account. With the increasing number of the channels and users, the state space and action space grow exponentially, leading to prohibitive computation complexity. In multi-user spectrum access scenario, the network can be modeled as a multi-agent system, which ought to optimize the system-level performance from the global view, rather than focus on maximizing single agent's utility. Multi-agent reinforcement learning (MARL) integrates the development of single-agent reinforcement learning, game theory and direct policy search techniques [19]. It is more promising to handle the large scale and complex wireless communication settings.

### *B. Contributions*

Our work is inspired by spectrum collaboration challenge (SC2), which was a three-year competition hosted by DARPA, aiming to overcome the scarcity of radio spectrum by machine learning approaches [20]. In the cognitive radio networks, there are two critical challenges, which are avoiding causing harmful interference to PUs and coordinating with other cognitive users to maximize the sum throughput of the system for cognitive users to access the licensed spectrum. We consider the cognitive radio network as a multi-agent system, and investigate the DSA of cognitive users by formulating it as a Markov game. Since there is no prior information about dynamics of the network, we cannot formulate the spectrum access model precisely by traditional approaches. Therefore, the concept of MARL is introduced and a QMIX algorithm [21] based multi-user DSA is proposed in multi-channel cognitive radio networks. QMIX is a novel value-based method which can train decentralized policies in a centralized end-to-end way. It employs a mixing network to estimate the joint-action value as a non-linear combination of per-agent values. In order to maximize the per-agent values while the joint-action value reaches maximum, the QMIX imposes a monotonicity constraint on the joint-action value with respect to the per-agent values, which guarantees the consistency between decentralized policies and centralized policy. The QMIX is accommodated for multi-agent cooperative settings, in which the agents achieve the optimal strategy by coordinating with each other. Our goal is to develop a distributed multi-agent learning architecture for multi-channel access under partial observability, and maximize the successful access rate of the whole system by designing suitable utility functions based on game theory. The main contributions of this paper are summarized as follows:

- It is a pioneer work that formulated the multi-channel access problem of multi-user as a Markov game. Specifically, we modeled the collaborative spectrum access as a cooperative game, and apply the concept of potential game to solve the distributed multi-user spectrum access problem.
- In order to tackle the instability and irrationality of the existing distributed access strategy, we proposed a centralized training and decentralized execution (CTDE) architecture, which estimates the joint-action value conditioning on global state information in the centralized training phase and the cognitive users make access decision condition only on their local partially observations. The CTDE architecture could be conducted with the advantages of emerging mobile edge computing technologies.

- We proposed a multi-user distributed spectrum access algorithm based on QMIX, also referred to as QMIX-DSA. With the proposed algorithm, each cognitive user can make access decisions based on its own local observation and requires no coordination with other cognitive users, which can greatly alleviate the communication overhead.

### *C. Organization of the paper*

The rest of the paper is structured as follows. We give some background knowledge about MARL, decentralized partially observable Markov decision progress (Dec-POMDP) and DRQN in section II. In section III, we present the system model and problem formulation. The proposed QMIX-DSA algorithm and the utility function design based on cooperative game are detailed in section IV. In section V, we implement the proposed algorithm and validate its effectiveness and efficiency by numerical results. Conclusions is drawn in section VI.

## II. Preliminary

### *A. Multi-agent reinforcement learning*

In multi-agent scenarios, a set of autonomous agents obtain optimal policy by interacting with their shared environment dynamically. Different from single-agent reinforcement learning in which the state transition of the environment switches based on its own action only, the new environment state of the multi-agent system is based on the joint-action of all agents. To a single agent specifically, its optimal policy depends not only on the state of the environment, but also on the policies of the other agents as well. Therefore, the shared learning environment is not stationary any more, and this can no longer be described by the Markov decision process (MDP) model. MARL is a multi-agent extension of the single-agent reinforcement learning and can be modeled as a Markov game [22]. Fig. 1 depicts a standard model of MARL.

MARL can be denoted as a tuple $\langle \mathcal{N}, \mathcal{S}, \{\mathcal{A}^n\}_{n\in\mathcal{N}}, \mathcal{P}, \{r^n\}_{n\in\mathcal{N}}, \gamma \rangle$, where $\mathcal{N} = \{1, 2, ..., N\}$ is the index set of agents, $\mathcal{S}$ denotes the set of environment states. $\mathcal{A}^n$ is the action space of agent $n$ and $\mathcal{A} = \mathcal{A}^1 \times \mathcal{A}^2 \times ... \times \mathcal{A}^n$ denotes the joint-action space of all agents, the joint-action vector is $\boldsymbol{a} = \{a^1, a^2, ..., a^n\}$. $\mathcal{P}(s'|s, \boldsymbol{a}) : \mathcal{S} \times \mathcal{A} \times \mathcal{S} \mapsto [0, 1]$ denotes the transition probability function of the system states given the joint-action $\boldsymbol{a}$. $r^n(s, \boldsymbol{a}) : \mathcal{S} \times \mathcal{A} \mapsto \mathbb{R}$ denotes the reward function of agent $n$ with the given joint-action $\boldsymbol{a}$ executed at state $s$. At each step, the $n$-th agent get the immediate reward $r^n(s, \boldsymbol{a})$ of a transition from $(s, \boldsymbol{a})$ to next state $s'$. In MARL, each agent aims to obtain an optimal policy $\pi^*$ which maximizes the expected cumulative reward during the executing period:

$$\pi^* = \arg\max_{\pi} E\left[\sum_{t=1}^{T} \gamma^{t-1} \cdot r_t^n(s, \boldsymbol{a}) \left| s_1 \right.\right], \tag{1}$$

where $0 \leq \gamma \leq 1$ is the discount factor, which indicates the long-term effect of the current action.

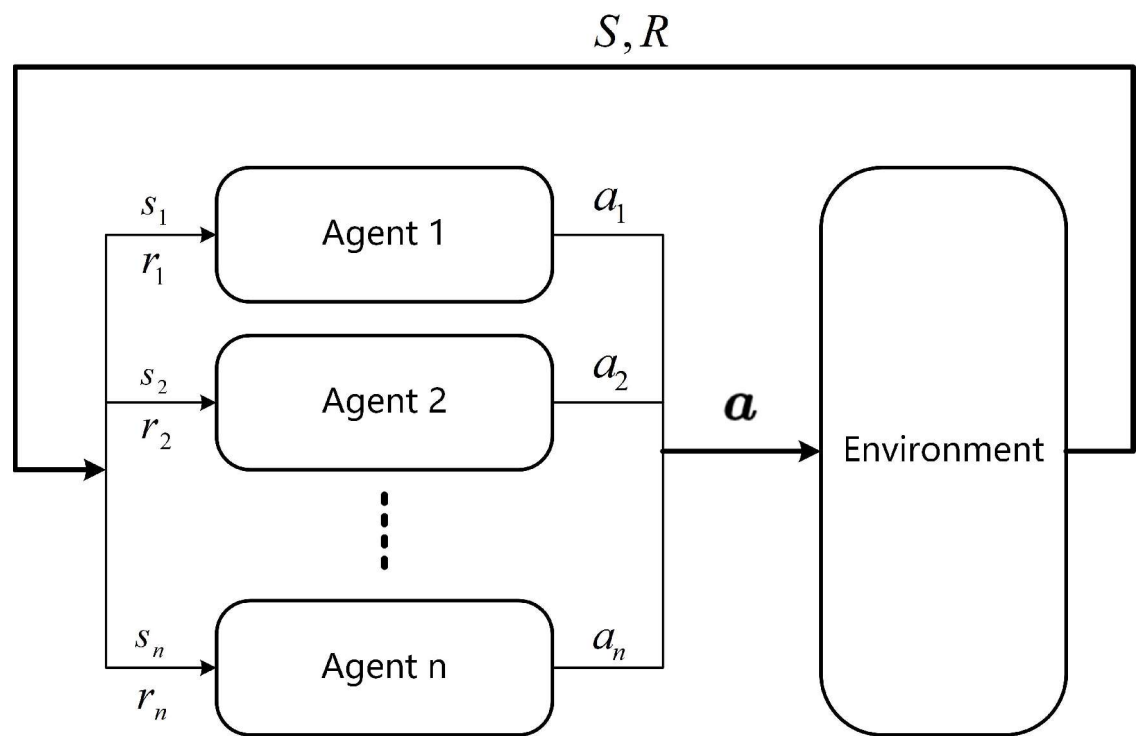

Fig. 1. The multi-agent reinforcement learning.

### *B. Decentralized partially observable Markov decision process*

MARL is applied in various areas such as autonomous navigation, traffic-light controlling, and sensor networks, where the agents are spatially distributed. Owning to the observation ability limitation of a single agent, an agent may not have full and perfect knowledge of the environment states, i.e., $z^n \neq s$, where $z^n$ is the observation and $s$ is the environment state. In single-agent settings, the problem is formulated as a POMDP. While in multi-agent scenarios, decisions are made in a decentralized manner by a set of agents, and the problem can be modeled as a Dec-POMDP [23], which offers a framework for cooperative sequential decision making in a shared environment. Dec-POMDP can be defined as a Markov game with incomplete information and denoted by a tuple $\langle \mathcal{N}, \mathcal{S}, \{\mathcal{A}^n\}_{n\in\mathcal{N}}, \mathcal{P}, \{r^n\}_{n\in\mathcal{N}}, \{\mathcal{Z}^n\}_{n\in\mathcal{N}}, \{\mathcal{O}^n\}_{n\in\mathcal{N}}, \gamma \rangle$. In contrast with the notations of MARL, $\mathcal{S}$ is the true state of the environment, $\mathcal{Z}$ is the observation space, and an observation function $\mathcal{O}(s, \boldsymbol{a}) : \mathcal{S} \times \mathcal{A} \mapsto \mathcal{Z}$ determines the observation that each agent observes at each step. The other denotations are the same as defined in MARL. Because the state is partially observed, it is beneficial to remember an observation-action history $\tau \in \mathcal{T} = \{\mathcal{Z} \times \mathcal{A}\}$. A stochastic policy $\pi(\boldsymbol{a}|\tau) : \mathcal{T} \times \mathcal{A} \mapsto [0, 1]$ is drawn conditioned on the observation-action history. Dec-POMDP is a stochastic game with incomplete information. At each step, action decisions of each agent are made solely on local observations, but these decisions can affect the global state and each agent's immediate reward and observation. The goal is to maximize the expected cumulative reward of each agent over a finite or infinite number of steps.

### C. Deep Recurrent Q-Network

POMDP is polynomial space-hard and requires exponential computation complexity to find the exact solution [24]. Even worse, the system dynamics is unknown in prior, which makes it intractable to find a feasible solution timely. In real-world MARL, the agents have only a partial view of the global state, resulting in imperfect awareness of the environment. It is difficult for normal DQN to address this kind of problem. To make up this gap, DRQN [25] introduces recurrent neural network (RNN) to deal with the partially observation problems, considering that RNN can remember a piece of historical data, and infer the environment state from the sequential but imperfect data. In DRQN architecture, a fully-connected layer is replaced by a recurrent layer (e.g., LSTM [26] or GRU [27]). In the training phase, the parameters of the recurrent layer and fully-connected layers are updated by back-propagation simultaneously. Many works have demonstrated that DRQN performs robustly with incomplete and imperfect information [14], [16].

## III. System model and problem formulation

### A. System model

In this paper, we investigate an OFDM-based cognitive radio network, which coexists with a primary network with $K$ orthogonal channels. The cognitive users access the spectrum holes opportunistically. The overall available frequency bandwidth is divided into $K$ orthogonal sub-channels with equal size, and the bandwidth spanned by an OFDM channel is smaller than the channel coherence bandwidth. Consequently, the spectrum can be approximated as flat. We consider the cognitive network as a fully synchronized time slotted system, where time is slotted and indexed by $t$, and each time slot has a fixed duration. In each time slot, $L$ OFDMA symbols are transmitted. The length of time slot is designed to match the coherence time of the channel, so the channel state information (CSI) holds the same within the time slot. Each cognitive user accesses the channels at the very beginning of the time slot, and holds on until the time slot ends. As depicted in Fig. 2, the system consists of primary tiers and cognitive tiers, in which the primary tiers consist of primary base stations and its associated PUs, whereas the cognitve tiers contains secondary access points (APs) and their associated cognitive users, or cognitive Tx-Rx pairs. The terms of cognitive users and SUs are interchangeable in this paper. The cognitive users are scattered within the transmission range of the primary BS, and all cognitive users locate in an area with geographical proximity and are of the sensing distance from each other (i.e., a fully-connected wireless network). We study the uplink spectrum access problem of cognitive users in a time-slotted channel hopping cognitive radio network, in which the scattered cognitive users transmit packets to their corresponding receivers or APs via shared wireless channels.

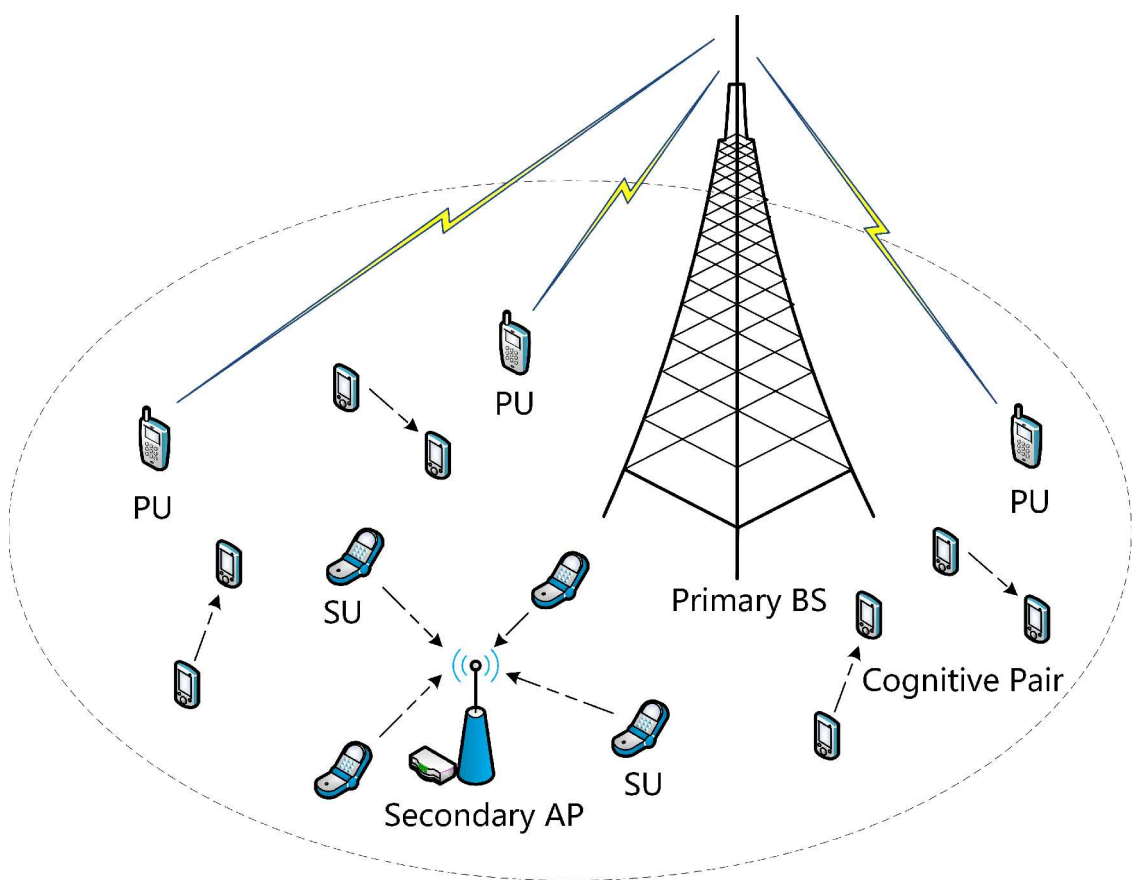

Fig. 2. The system model.

Without loss of generality, we assume that the cognitive radio network consists of $N$ cognitive users, and $n \in \{1, 2, ..., N\}$ is the cognitive users index. We focus on the subchannels access in the uplink of cognitive radio network, in which $N$ cognitive users share $K$ orthogonal channels, and $k \in \{1, 2, ..., K\}$ denotes the $k$-th subchannel. The cognitive users transmit the data over the shared channels cooperatively using the random-access protocol (e.g., CSMA-type narrowband transport protocol). At each time slot, each cognitive user selects $M \, (0 \leq M \leq K)$ channels out of $K$ orthogonal channels to sense, and then chooses one channel from the sensed channels to transmit packets according to its learnt policy. We assume that all cognitive users are backlogged and always attempt to transmit at each time slot.

We consider the distributed DSA problem of multiple cognitive users. To guarantee the scalability and robustness of the system, we assume that there is no central controller in the cognitive radio network. In order to provide centralized off-line training, we deploy a parameter server on the edge of the cognitive radio network (i.e., wireless APs and cognitive base station). The parameter server is specially deployed for training, and is not involved into making spectrum access decisions. In order to enhance the efficiency of spectrum access protocol, we assume that there is no information exchange between cognitive users. Each cognitive user makes the access decision according to its own learnt policy based on local observation.

### *B. Problem formulation*

Based on the system model we proposed above, the distributed multi-user DSA problem can be modeled as a cooperative Markov game [28]. Considering that cognitive users have limited sensing ability and no prior knowledge of the network dynamics, the cooperative spectrum access can be modeled as a Dec-POMDP, which is given by a tuple $\mathcal{G} = \langle \mathcal{N}, \mathcal{S}, \{\mathcal{A}^n\}_{n\in\mathcal{N}}, \mathcal{P}, \{r^n\}_{n\in\mathcal{N}}, \{\mathcal{Z}^n\}_{n\in\mathcal{N}}, \{\mathcal{O}^n\}_{n\in\mathcal{N}}, \gamma\rangle$. $\mathcal{N}$ is the number of cognitive users, $\mathcal{S}$ is the true state of overall channels occupation. $\mathcal{A}$ is the channel-selection space of cognitive users which yields the joint-action set $\mathcal{A} = \mathcal{A}^1 \times \mathcal{A}^2 \times ... \times \mathcal{A}^n$ and joint-action vector $\boldsymbol{a} = \{a^1, a^2, ..., a^n\}$. $\mathcal{P}(s'|s, \boldsymbol{a}) : \mathcal{S} \times \mathcal{A} \times \mathcal{S} \mapsto [0, 1]$ is the transition probability function of the channels states given the joint-action $\boldsymbol{a}$, $r^n(s, \boldsymbol{a}) : \mathcal{S} \times \mathcal{A} \mapsto \mathbb{R}$ denotes the reward function of the cognitive users with the given joint-action $\boldsymbol{a}$ executed at current state $s$. $\mathcal{Z}^n$ is the observation space, and an observation function $\mathcal{O}^n(s, a) : \mathcal{S} \times \mathcal{A} \mapsto \mathcal{Z}$ determines the observation that each cognitive user sensed at each step. $\gamma \in [0, 1]$ is the discount factor, which indicates the long-term effect of the current action. Considering the state is partially observed and imperfect, we exploit the observation-action history $\tau \in \mathcal{T} = \{\mathcal{Z} \times \mathcal{A}\}$ to infer the true state from the sequential but incomplete data. Let $\mathcal{B}(\tau) : \mathcal{T} \mapsto \mathcal{S}$ be the belief function maintained by the cognitive users. We employ DRQN to update $\mathcal{B}(\tau)$ in each time slot, and infer the true channels occupation state. So, the cognitive users can make access decisions according to the learnt policy solely based on their own local observations.

*1) State switching patterns:* In the cognitive radio network with multi-user sharing available channels cooperatively, each cognitive user first selects $M\,(0 \leq M \leq K)$ channels out of $K$ channels to sense at each time slot, and then chooses one idle channel to access. The channels may be independent or correlated. For each channel, it has two states: *idle* $(1)$ or *busy* $(0)$, and its state transition follows two-state Markov chain, denoted as $\mathcal{P}_k = \begin{bmatrix} p_{00} & p_{01} \\ p_{10} & p_{11} \end{bmatrix}$, as depicted in Fig. 3. Consequently, the whole system can be described as a $2^K$-state Markov chain. We denoted the channel states as $s_t = \{s_{1,t}, s_{2,t}, ..., s_{K,t}\}$, where $s_{k,t}$ stands for the state of the $k$-th channel in time slot $t$. The channels state switch at the beginning of each time slot and remain the same within the time slot duration.

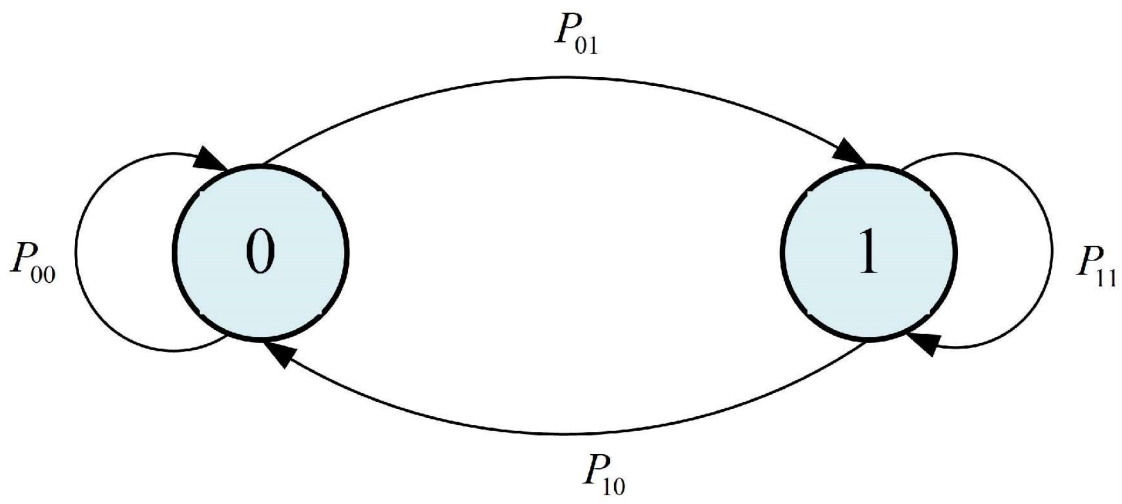

Fig. 3. Two-state Markov model of channel occupancy.

The cognitive users adot the Listen-Before-Talk (LBT) mechanism to access the spectrum. At the very beginning of each time slot, each cognitive user selects $M$ channels to sense. If there is more than one idle channel in the sensed channels, the cognitive user chooses one idle channel to transmit packets. At the end of the time slot, if the cognitive user receives an ACK from the corresponding receiver, the transmission is successful, and a collision occurs otherwise. The slot structure is depicted in Fig. 4.

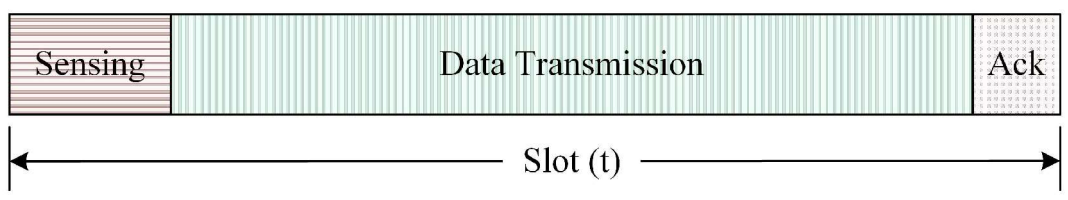

Fig. 4. The slot structure.

*2) Cognitive users' observation space:* Considering the network dynamics is not priorly known to the cognitive users, and the perception of the cognitive users is limited, the cognitive users only have partially local information of the channels state. We modeled the DSA of the multi-user as a Dec-POMDP, and we can deduce the state switching pattern from an observation-action history $\tau \in \mathcal{T} = \{\mathcal{Z} \times \mathcal{A}\}$. At each time slot, each cognitive user selects $M$ channels out of the $K$ channels to sense, and the states of the selected channels are revealed to the cognitive user, while the states of the other channels which are not selected remain unknown. The size of the observation space $\mathcal{Z}$ is $2^M \times \begin{pmatrix} K \\ M \end{pmatrix}$. We denote the observation of cognitive user $n$ as $z^n(t) = \{z^n_{1,t}, z^n_{2,t}, ..., z^n_{K,t}\}$, where $z^n_{k,t}$ stands for the observed state of $k$-th channel by the $n$-th cognitive user in time slot $t$. $z^n_{k,t}$ is defined as

$$z^n_{k,t} = \delta^n_{k,t} s_{k,t} = \begin{cases} s_{k,t}, & \text{if the } k\text{-th channel is sensed by } n\text{-th cognitive user in time slot } t, \\ -1, & \text{if the } k\text{-th channel is not sensed by } n\text{-th cognitive user in time slot } t. \end{cases} \tag{2}$$

$s_{k,t}$ is the state of the $k$-th channel in time slot $t$, which is defined as

$$s_{k,t} = \begin{cases} 1, & \text{if the } k\text{-th channel is } \textit{idle} \text{ in time slot } t, \\ 0, & \text{if the } k\text{-th channel is } \textit{busy} \text{ in time slot } t. \end{cases} \tag{3}$$

And $\delta^n_{k,t}$ is an indicator of whether the channel is sensed, that is defined as

$$\delta^n_{k,t} = \begin{cases} 1, & \text{if the } k\text{-th channel is sensed by } n\text{-th cognitive user in time slot } t, \\ 0, & \text{if the } k\text{-th channel is not sensed by } n\text{-th cognitive user in time slot } t. \end{cases} \tag{4}$$

*3) Cognitive users' action space:* As described before, each cognitive user selects $M$ channels out of the $K$ channels to sense in each time slot, where $1 \leq M \leq K$. Therefore, the size of the action space $\mathcal{A}$ is $\begin{pmatrix} K \\ M \end{pmatrix}$. In this paper, the cognitive users are homogeneous, and they have the same observation space and action space. At each time slot, each cognitive user executes one action selected from the action space according to the learnt policy, and acquires the state of the sensed channels.

*4) Cooperative game reward:* At each time slot, after sensing the $M$ selected channels, the cognitive users learn the state of the sensed channels. If there is no idle channel in the sensed channels, the cognitive user does not transmit data and receives reward $(0)$ at the end of the time slot. If there are idle channels in the sensed channels, the cognitive user picks one idle channel to transmit packets and receives an ACK from the corresponding receiver at the end of the time slot. Through measuring the Signal-to-Interference-plus-Noise Ratio (SINR) of the feedback from the corresponding receiver, we can infer whether there is a collision. If the transmission is successful, the cognitive user acquires a positive reward $(+2)$. If a collision occurs, the cognitive user acquires a negative reward $(-1)$. The reward acquired by each cognitive user is defined as

$$r_t^n = \begin{cases} 0, & \begin{array}{l}\text{if the } n\text{-th cognitive user does not transmit}\\ \text{in time slot } t,\end{array} \\ +2, & \begin{array}{l}\text{if the } n\text{-th cognitive user transmits and}\\ \text{no collisions occur in time slot } t,\end{array} \\ -1, & \begin{array}{l}\text{if the } n\text{-th cognitive user transmits and}\\ \text{collisions occur in time slot } t.\end{array} \end{cases} \tag{5}$$

From the perspective of one single cognitive user, the goal is to learn an individual multi-channel access strategy $\pi^*$, which maximizes the expected cumulative reward during the executing period $T$, also named an episode. Therefore, we have

$$\pi^* = \arg\max_{\pi} E\left[\sum_{t=1}^{T} \gamma^{t-1} \cdot r_t^n(z^n, a^n)\,|z_1^n\right], \tag{6}$$

where $0 \leq \gamma \leq 1$ is the discount factor, which indicates the long-term effect of the current action. In this paper, owning to episodes are bounded and each time slot is equivalent, we set $\gamma = 1$. When we consider an episode composing only one time slot, the problem is simplified to the well-known multi-armed bandit problem.

In a multi-user scenario, where a group of cognitive users share the common available channels, the number of cognitive users may be much more than that of channels, i.e., $N > K$. Therefore, the cognitive users should coordinate to access the spectrum holes cooperatively in order to reduce collisions so as to maximize the throughput of the cognitive radio network. To address this problem, we model the distributed DSA problem as a cooperative game [29], in which all cognitive users share the global system-level reward. The global reward is defined as

$$R^{tot} = \sum_{n=1}^{N} \sum_{t=1}^{T} \gamma^{t-1} \cdot r_t^n(z^n, \boldsymbol{a}). \tag{7}$$

As defined above, $r_t^n$ is a linear function, and the global reward can be rewritten as

$$R^{tot} = \sum_{t=1}^{T} \gamma^{t-1} \sum_{n=1}^{N} r_t^n(z^n, \boldsymbol{a}). \tag{8}$$

In DRL, the global reward is received at the end of the episode. Thus, it is a delayed reward, and has low training efficiency. From Equ. (8), we can define the total reward of all cognitive users in each time slot as

$$r_t^{tot} = \sum_{n=1}^{N} r_t^n(z^n, \boldsymbol{a}). \tag{9}$$

The multi-user spectrum access in each time slot can also be formulated as a cooperative game. Cooperative games are a special type of exact potential game (EPG) [30], and $r_t^{tot} : \mathcal{Z} \times \mathcal{A} \mapsto \mathbb{R}$ is the potential function. Denote $a^n, \widetilde{a}^n \in \mathcal{A}$ as the chosen action of the $n$-th cognitive user, and $a^{-n} = \{a^i\}_{i \neq n}$ as the action profile of the other cognitive users except $n$-th cognitive user. Then the following equation holds for all cognitive users:

$$r_t^n(a^n, a^{-n}) - r_t^n(\widetilde{a}^n, a^{-n}) = r_t^{tot}(a^n, a^{-n}) - r_t^{tot}(\widetilde{a}^n, a^{-n}). \tag{10}$$

Therefore, $r_t^{tot}$ has the same monotonicity as $r_t^n$, which is equivalent to

$$\arg\max_{\boldsymbol{a}} \ r_t^{tot}(s, \boldsymbol{a}) = \begin{pmatrix} \arg\max_{a^1} r_t^1(z^1, a^1) \\ \vdots \\ \arg\max_{a^N} r_t^N(z^N, a^N) \end{pmatrix}. \tag{11}$$

EPG admits an important property that it has the finite improvement property (FIP) and convergence to a pure-strategy Nash equilibrium (NE) when the strategy space is finite [30]. The goal of distributed DSA of multi-user is to find the equilibrium point that maximizes the expected cumulative reward

$$G^{tot} = \max_{\pi^*} E\left[\sum_{t=1}^{T} \gamma^{t-1} \sum_{n=1}^{N} r_t^n(z^n, a^n, a^{-n}) | s_1\right]. \tag{12}$$

In the following section, we will employ MARL to get the near-optimal strategy of this Markov game in an incomplete and dynamic environment.

## IV. The proposed QMIX-based dynamic spectrum access algorithm

### A. *Centralized training and decentralized execution*

The existing works mainly focus on single-agent reinforcement learning or centralized approaches, where a network controller is employed as a centralized trainer and executor. However as the number of users and network scale increases, the state space and action space increase exponentially, which brings prohibitive computational overhead. MARL is capable to model and resolve the multi-agent system problem in a distributed manner, even when the cognitive users only observe their local information.

In practice, there is usually no base station or central controller in cognitive radio networks. Even though there is a central controller, the exchange of observation and decision-making information between central controller and cognitive users may impose heavy communication overhead on the network. Meanwhile, if the cognitive radio network is fully decentralized, and there is no coordination information exchange between cognitive users, the system would be nonstationary and nonconvergent. To make a tradeoff between the aforementioned two extremes, we propose a centralized training and distributed execution architecture, which employs a centralized trainer on the edge of the cognitive radio network, as shown in Fig. 5. The CTDE architecture is robust to network topology variations and facilitates scalability.

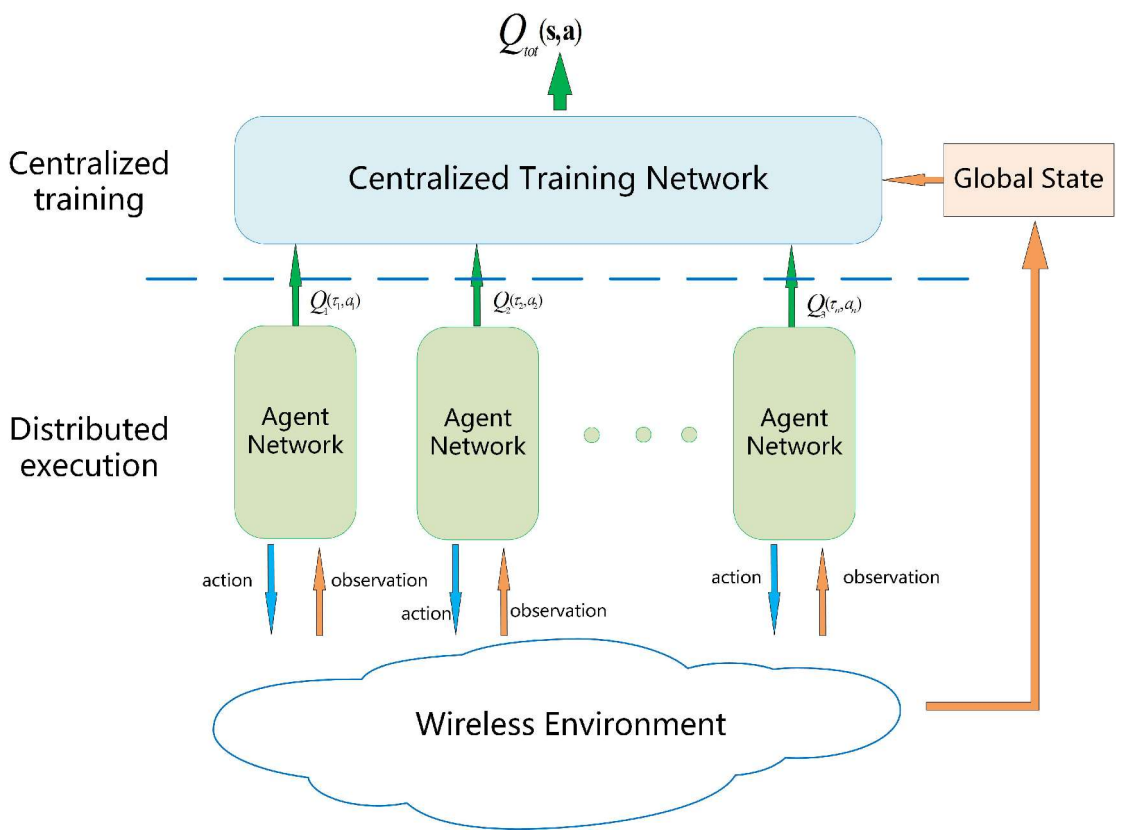

Fig. 5. The centralized training and distributed execution architecture.

In the training phase, cognitive users first collect their observations, actions and rewards and then send them to the centralized trainer which is deployed on the edge computing server. In the execution phase, cognitive users make spectrum access decisions on their own autonomously according to their local observations.

### *B. The proposed QMIX-DSA algorithm*

For MARL with discrete state and action space, the value-based algorithms are effective [8]. Value decomposition network (VDN) [31] is a centralized learning and distributed execution algorithm, which is accommodate for solving cooperative games. VDN obtains the joint-action value function by adding all agents' Q-value function directly, i.e., $Q^{tot} = \sum_{n=1}^{N} Q^n(\tau^n, a^n; \theta^n)$, where $Q^{tot}$ stands for joint-action value function, $Q^n(\tau^n, a^n; \theta^n)$ denotes the $n$-th agent's Q-value function, $\tau = (\tau^1, \tau^2, ..., \tau^N)$ denotes the joint observation-action history and $\tau_t^n = (z_{t-l}^n, a_{t-l}^n, ..., z_{t-1}^n, a_{t-1}^n, z_t^n, a_t^n)$ denotes the $l$-length observation-action history of $n$-th agent until step $t$, and $\boldsymbol{a} = (a^1, a^2, ..., a^N)$ denotes the joint-action. $\theta^n$ is the parameters of $n$-th agent network. Considering maximizing $Q^{tot}$ is equivalent to maximizing all $Q^n(\tau^n, a^n; \theta^n)$, we can obtain the distributed policy for all the agents through maximizing the $Q^n(\tau^n, a^n; \theta^n)$ respectively. QMIX [21] is an extension of the VDN, which adopts a mixing network to integrate all individual Q-value functions of the agents, meanwhile coalesces the global state information to assist training, which is helpful to improve the performance and convergence speed of the algorithm. QMIX is also a centralized learning and distributed execution reinforcement learning algorithm, whose $Q^{tot}$ has the same monotonicity with the single agent value function $Q^n(\tau^n, a^n; \theta^n)$. i.e.,

$$\arg\max_{\boldsymbol{a}} Q^{tot}(\tau, \boldsymbol{a}) = \begin{pmatrix} \arg\max_{a^1} Q^1(\tau^1, a^1) \\ \vdots \\ \arg\max_{a^N} Q^N(\tau^N, a^N) \end{pmatrix}. \tag{13}$$

The monotonicity can be enforced through imposing a constraint on the relationship between $Q^{tot}$ and each $Q^n$, that is

$$\frac{\partial Q^{tot}}{\partial Q^n} \geq 0, \forall n \in \{1, 2, ..., N\}. \tag{14}$$

Consequently, in the cooperative game, the common joint-action value function is the potential function. Maximizing $Q^{tot}$ is equivalent to maximizing all $Q^n(\tau^n, a^n; \theta^n)$, and thus we can obtain the distributed policy for all the agents by maximizing $Q^n(\tau^n, a^n; \theta^n)$ respectively. QMIX consists of a mixing network, a hyper-network [32] and a set of agent networks. Each agent has a DRQN which receives the current local observation and the last action as input at each step, and estimates its individual Q-value function $Q^n(\tau^n, a^n; \theta^n)$. The mixing network takes the agent networks' outputs as input and mixes them monotonically to produce the global value function $Q^{tot}$. The weights of the mixing network are produced by hyper-network, which takes the global state as input and generates the non-negative weights for the mixing network to guarantee the monotonous of $Q^{tot}$. QMIX is accordance with the CTDE architecture as depicted in Fig. 5.

In cognitive radio networks, we propose the QMIX-DSA algorithm based on QMIX, which formulates the multi-user DSA problem as a cooperative game. Since the cognitive users are

TABLE I

QMIX-DSA ALGORITHM

```
QMIX-DSA Algorithm: Training Phase
Initialize the wireless environment and experience replay buffer D;
Initialize the parameters of hypernetwork and N agent DRQN networks θ;
Set the target-network parameters θ⁻ = θ; set the learning rate α;
while epoch ≤ epoch max do
  for episode ≤ episode max in each epoch do
    for slot t ≤ slot length T in each episode do
      for each cognitive user n do
        Get each cognitive user's action a_t and each user's RNN hidden states;
      end
      Get reward r_t and next observation z_{t+1};
      Store the {z_t, a_t, r_t, z_{t+1}} to observation-action history;
      slot = slot + 1;
    end
    Store the episode to replay buffer D;
    episode = episode + 1;
  end
  for train ≤ train step max in each epoch do
    Sample a batch of B episodes from replay buffer D;
    for each slot in each episode in sampled batch do
      Get Q_t from the evaluate-network;
      Get Q_{t+1} from the target-network;
      Update each agent's RNN hidden state using transitions from batch;
    end
    Calculate the loss function
    L(θ) = (1/T)(1/B) Σ_n Σ_t ((r_t+γQ_{t+1}) − Q_t)^2;
    Update the evaluate-network parameters θ=θ−α∇_θ L(θ);
    if update-interval train then
      Update the target-network parameters θ⁻ = θ;
    end
  end
end
```

homogeneous, their DRQN networks are identical and have the same parameters. The QMIX-DSA algorithm is detailed in Table I.

In the execution phase, cognitive users make spectrum access decisions individually according to the learnt policy based on their own local observations.

## V. Experiments and performance evaluations

In this section, we present numerical experiments to validate the effectiveness and efficiency of the proposed QMIX-DSA algorithm. Specifically, the proposed algorithm is compared with the state-of-the-art DQSA algorithm, which achieves the best performance among existing works tackling the DSA problem of multi-user in multi-channel cognitive radio networks. We consider a cognitive radio network with $K = 16$ orthogonal channels whose bandwidths are equal, and each channel is occupied by corresponding PUs following the two-state Markov chain pattern. $N$ cognitive users scattering in proximity access the spectrum holes cooperatively as described in Section III. Here, we consider various scenarios and settings to fully validate the proposed QMIX-DSA algorithm. First, we validate the QMIX-DSA algorithm in the scenarios both $N \leq K$ and $N > K$ (i.e., the numbers of cognitive users $N$ varies from 3 to 21). Then, we present the numerical results when cognitive users possess different numbers of sensing channels. Further on, in order to evaluate the robust of proposed algorithms, we conduct experiments in periodic channels setting, correlated channels setting and variable wireless environment. At last, we validate the effectiveness of the proposed QMIX-DSA algorithm in real-world channels setting.

The QMIX-DSA algorithm is comprised of a mixing network, a hyper-network and $N$ agent networks. The mixing network consists of one hidden layer of 32 neurons with ELU as the activation function. The hypernetwork which produces non-negative weights for mixing network also consists of one hidden layer of 32 neuros with ReLU as the activation function. Each agent network is a DRQN with one recurrent layer consisting of a GRU with 64-dimension hidden state. In the training phase, each cognitive user adopts $\varepsilon$-greedy policy to select action. $\varepsilon$ decays from 0.4 to 0.05 over 10000 steps and keeps constant until the end. The discount factor $\gamma = 1$ and learning rate $\alpha = 5e - 4$. Every time we sample batches of 16 episodes from the replay buffer, and each episode contains 20 time slots. The target networks are updated every 40 training steps. In order to speed up the training, all cognitive users share the same agent network architecture.

### A. *Varied numbers of cognitive users with fixed number of sensing channels*

In this experiment setting, 16 orthogonal channels are occupied by PUs following distinct two-state Markov chain patterns $\mathcal{P}^k$. We randomly set the state switching probabilities $\mathcal{P}^k = [(x, 1-x), (y, 1-y)]$ for $k$-th channel. $x$ denotes the probability of the state transferring from *busy* to *idle*, and $1-x$ denotes the probability of state staying *busy*. $y$ denotes the probability of the state transferring from *idle* to *busy*, and $1-y$ denotes the probability of the state staying *idle*. We conduct the experiments for $N=3$, 6, 9, 15 and 21 cognitive users, and each cognitive user percieves two channels at each time slot. Fig. 6 demonstrates that the proposed QMIX-DSA algorithm can converge with low collision rate after certain training steps for different numbers of cognitive users. However, when the number of cognitive users is almost equal to or more than the number of channels, the convergence speed of the proposed QMIX-DSA algorithm slows down. This is due to the action space becomes larger as the number of cognitive users increasing. The outcome shows that the proposed algorithm has good scalability.

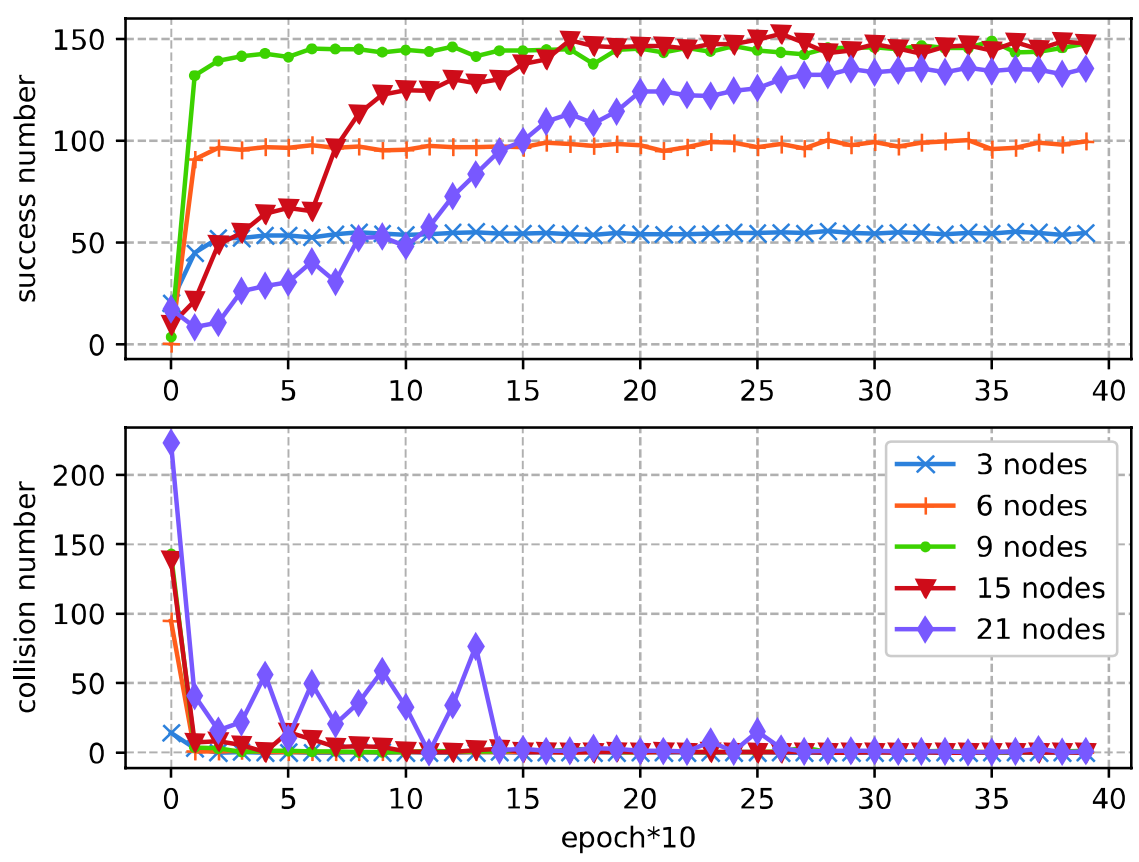

Fig. 6. Varied numbers of cognitive users with two sensing channels.

### *B. Fixed number of cognitive users with varied numbers of sensing channels*

The experiment setting is the same as subsection $A$, and we implement the experiments for three cognitive users with varied sensing channels $M=1$, 2, 3 and 4. By statistical calculation, 16 orthogonal channels have 7.42 idle channels in each time slot on average. Each episode is comprised of 20 time slots, and thus each episode can release 148.4 idle time slots for reusing. For three cognitive users with continuous transmission, they need 60 idle time slots for accessing in an episode. From Fig. 7, we can observe that the proposed algorithm converges faster and achieves better performance when cognitive users perceive more channels, meanwhile causes the lower collision rate. For three cognitive users with four and five sensing channels, they access almost 57 idle time slots in an episode, and the success rate of access is more than 95%.

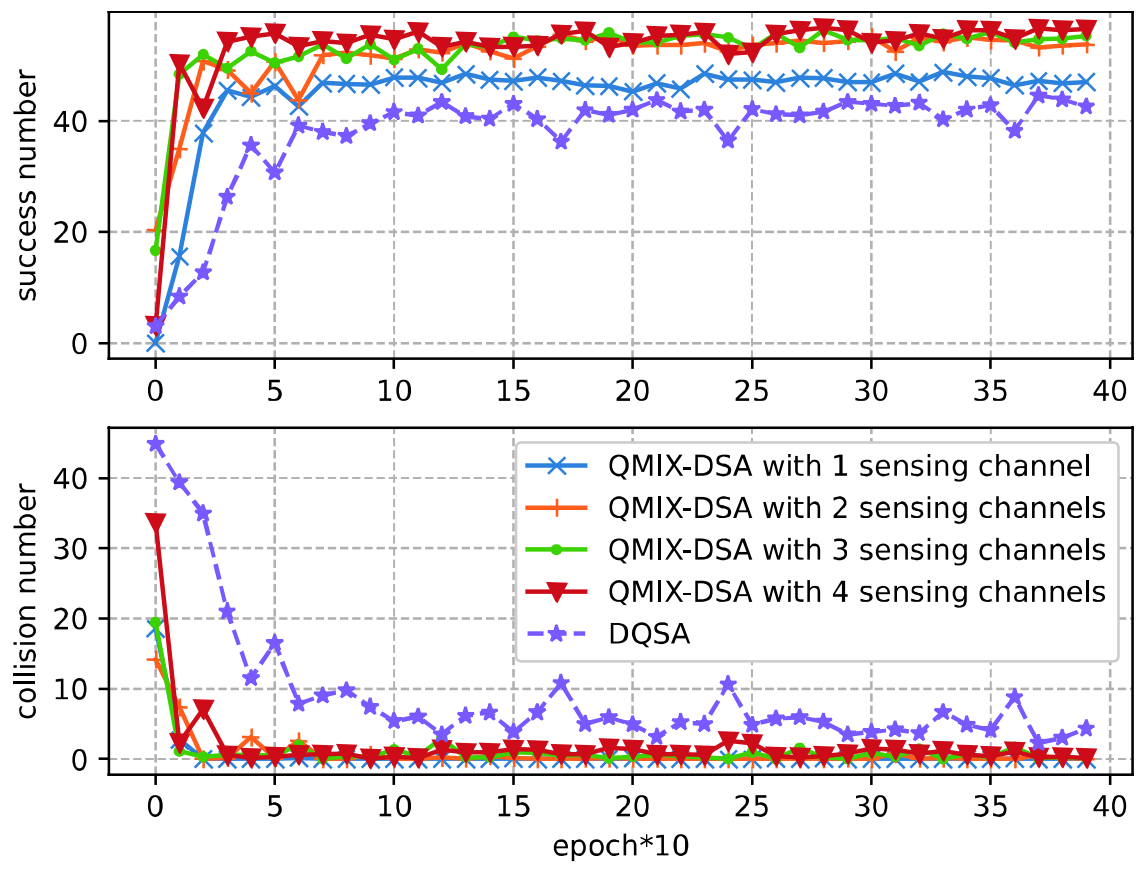

Fig. 7. Three cognitive users with varied numbers of sensing channels vs DQSA.

In the experiments of six cognitive users with varied sensing channels $M = 1$, 2, 3 and 4, we can get the same conclusion as scenarios with three cognitive users. For six cognitive users with continuous transmission, they need 120 idle time slots for accessing in an episode. As depicted in Fig. 8, six cognitive users with four sensing channels access almost 112 idle time slots in an episode, the success rate of access is more than 92%.

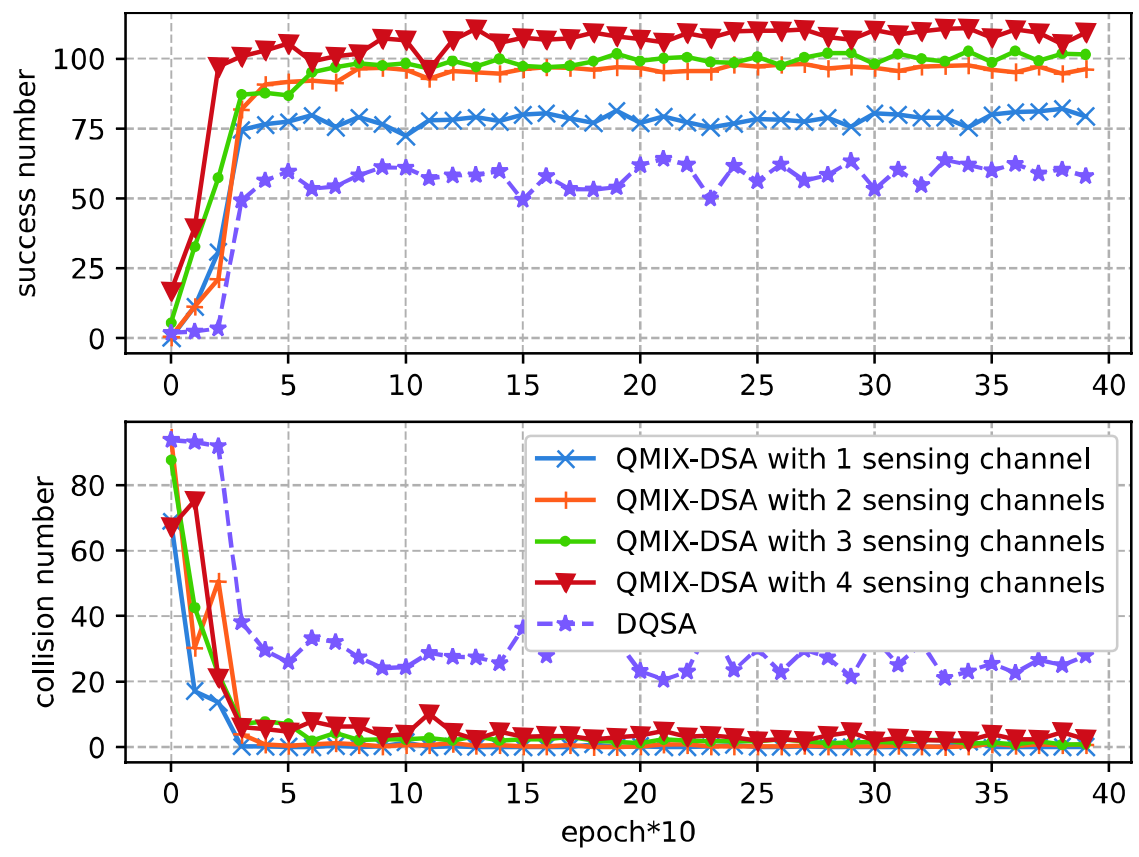

Fig. 8. Six cognitive users with varied numbers of sensing channels vs DQSA.

Furthermore, we implement the DQSA algorithm proposed by [16], in which the authors devised a deep Q-learning spectrum access algorithm based on Dueling-DQN and LSTM under partial observability. Each cognitive user employs an independent single-agent Q-Learning network, and the central trainer is responsible for updating the network parameters. The DQSA algorithm also formulates the distributed DSA problem as a cooperative game in the centralized training phase, and achieves almost 80% channel throughput in the open shared radio access environment. From Fig. 7, we observe that three cognitive users employing DQSA reuse about 40 idle time slots in an episode with 20 time slots, and the convergence curve fluctuates wildly. The success rate of access is 67% with PUs coexisting. From Fig. 8, we observe that six cognitive users employ DQSA reuse about 62 idle time slots in an episode and the collisions increase. The success rate of access is 52%. As the number of cognitive users increasing, the successful access rate decrease dramatically.

From the experiments conducted above, three cognitive users with four sensing channels and six cognitive users with two sensing channels achieve almost the optimal performance. This is because at centralized training phase, we formulate the distributed spectrum access as a cooperative game, and cognitive users learn a coordinate policy in a centralized training fashion. As each cognitive user perceive more channels, they can obtain the complete globe state information at centralized training phase. We also observe that the QMIX-DSA algorithm outperforms the state-of-the-art DQSA, especially when the number of cognitive users is large. This is due to the DQSA cannot lean a coordinate policy as the number of cognitive users increasing.

### *C. Periodic channels*

In this experiment, we set four idle channels in each time slot, and the channels with idle slot switch according to round-robin scheduling, the switching probability is set to 0.75. The channel switching pattern is depicted in Fig. 9.

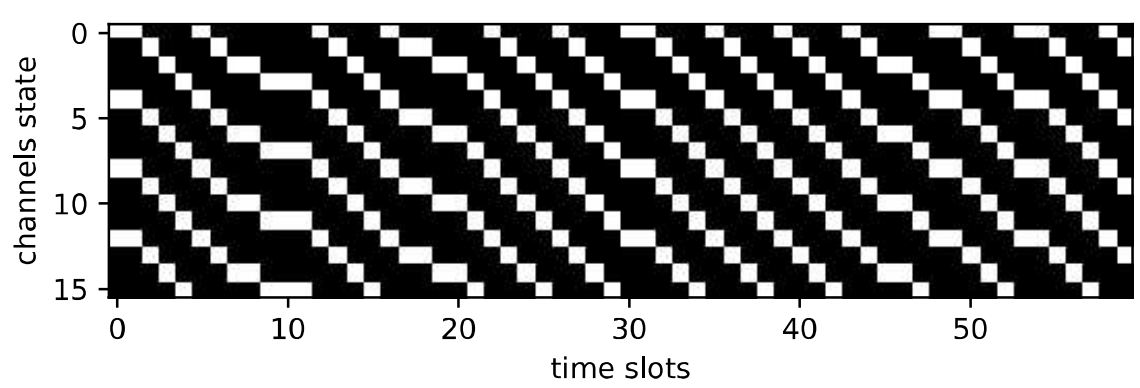

Fig. 9. Periodic channels with four idle channels in each time slot, and the switching probability is 0.75.

In this scenario, we implement the QMIX-DSA algorithm for three cognitive users and each user with two sensing channels. In an episode comprised of 20 time slots, three cognitive users with continuous transmission need 60 idle time slots to access. From Fig. 10, we can observe that the proposed QMIX-DSA algorithm converges after certain training steps, and achieves the near-optimal performance. This implies that the cognitive users have leant the channels switching pattern and adopt a cooperative access policy.

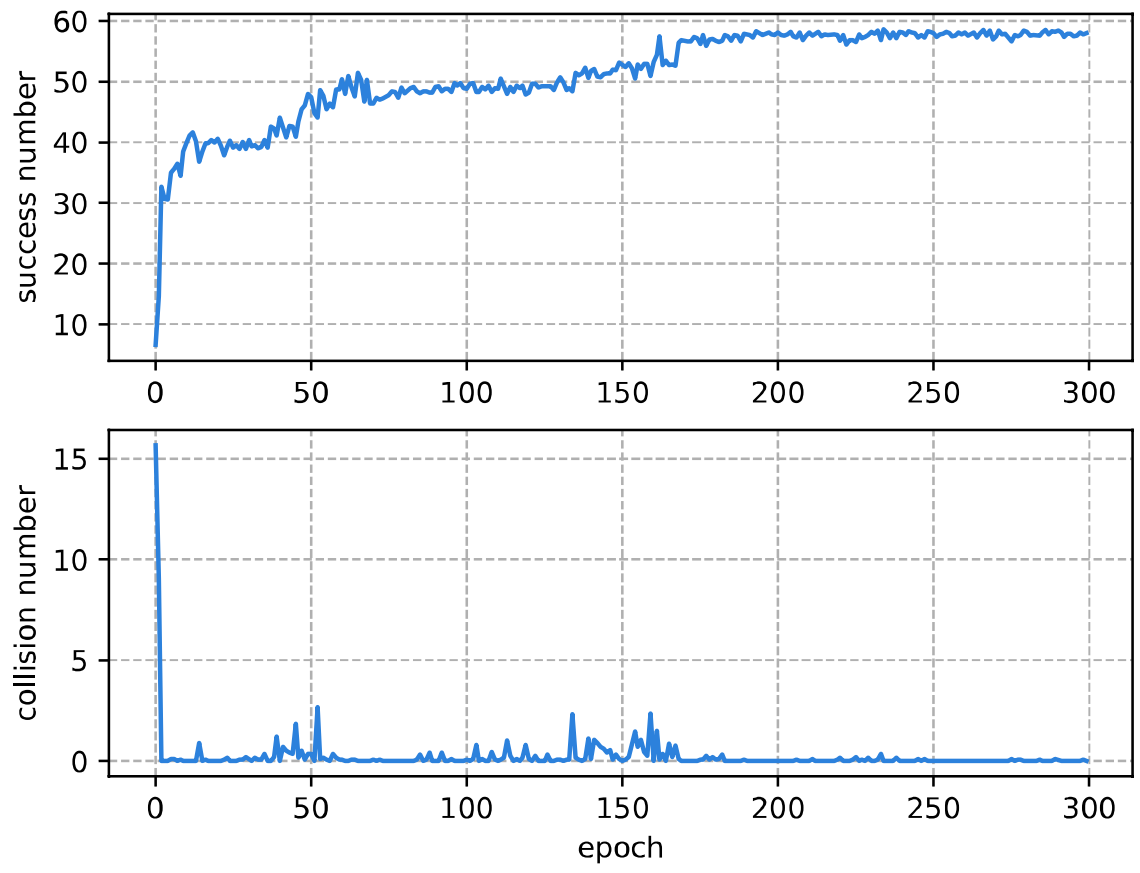

Fig. 10. Three cognitive users with two sensing channels in periodic channels.

## *D. Correlated channels*

In this experiment, we consider wireless channels with switching pattern of correlated channels. 16 orthogonal channels are divided into four subsets, each channel in a subset is the same or inverse to the other channels. By statistical calculation, 16 orthogonal channels have 8 idle channels in each slot on average. The channel switching pattern is depicted in Fig. 11.

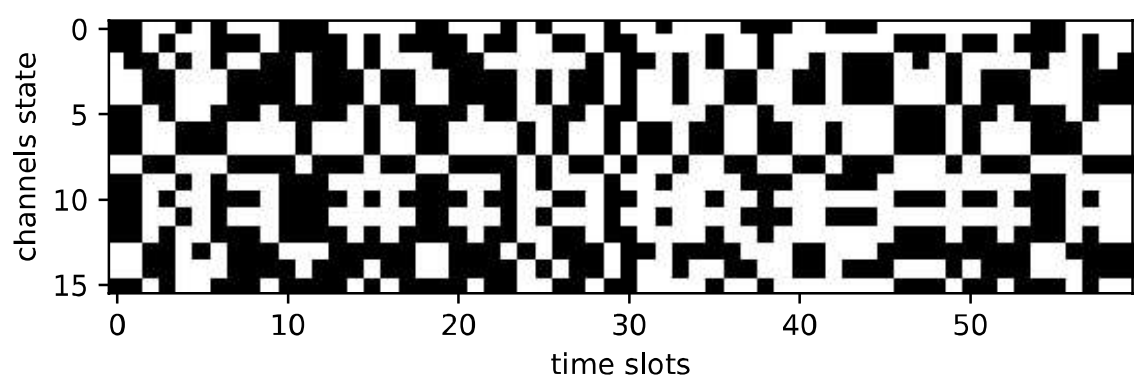

Fig. 11. Correlated channels with four set, the size of each set may be different.

In this scenario, we implement the QMIX-DSA algorithm for three cognitive users with two sensing channels in correlated channels. In an episode comprised of 20 time slots, three cognitive users with continuous transmission need 60 idle time slots to access. From Fig. 12, we observe that the proposed algorithm converges after certain training steps, and achieves the near-optimal performance. This implies that the cooperative access policy is learnt in correlated channels setting.

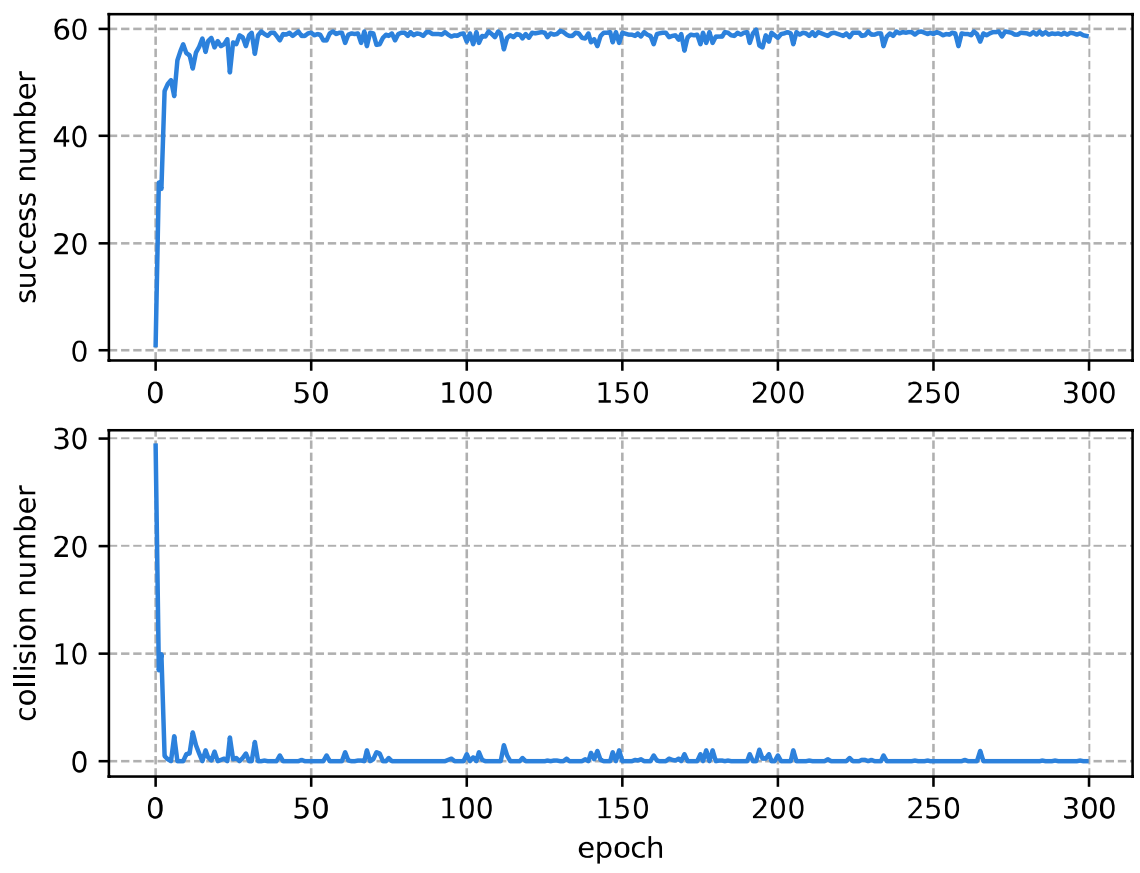

Fig. 12. Three cognitive users with two sensing channels in correlated channels.

### *E. Adaptivity to variable environments*

In order to illustrate the adaptivity of the proposed framework, we implement the proposed QMIX-DSA algorithm in the time-varying environments.

*1) Periodic channels to correlated channels:* In this experiment, we first set the wireless channels as a round-robin switching pattern, which is described in subsection *C*. Then at epoch 150, the channels switching pattern changes to correlated channels which is described in subsection *D*. The time point when to change the switching pattern is unknown to the cognitive users. In the execution phase, when the cognitive users encounter a sudden performance degradation, this means that the wireless environment is changed. Then, all parameters of the networks are reset to initial states, and a new training phase initiates. In the described senarios, we implement the QMIX-DSA algorithm for three cognitive users with two sensing channels. From Fig. 13, we observe that the proposed algorithm can converge fast and achieve the near-optimal performance, when the cognitive users encounter a sudden performance degradation at epoch 150.

*2) Markov channels to Markov channels:* In this experiment, we first set the wireless channels following Markov switching pattern $\mathcal{P}$, which is depicted in subsection $A$. Then at epoch 150, the channels switching pattern changes to $\mathcal{P}'$, and $\mathcal{P}' \neq \mathcal{P}$. The time point when to change the switching pattern is unknown to the cognitive users. As described above, when the cognitive users encounter a sudden performance degradation in the execution phase, all parameters of the networks are reset, and to initiate a new training phase. We implement the QMIX-DSA algorithm for three cognitive users with two sensing channels in this dynamic environment. From Fig. 14, we observe that the proposed algorithm can converge fast and achieve the near-optimal performance, when the cognitive users encounter a sudden performance degradation at epoch 150. This implies that the proposed QMIX-DSA algorithm can be applied to dynamic scenarios.

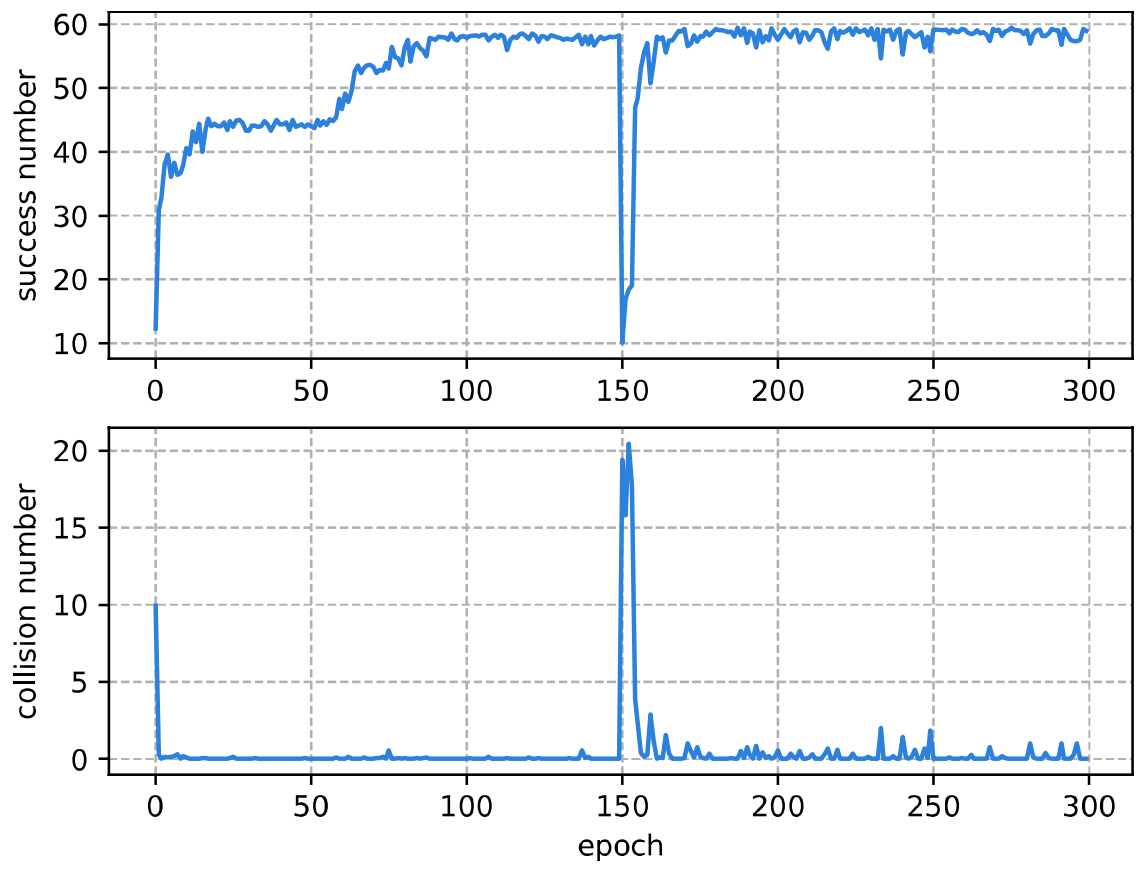

Fig. 13. Three cognitive users with two sensing channels from periodic channels to correlated channels.

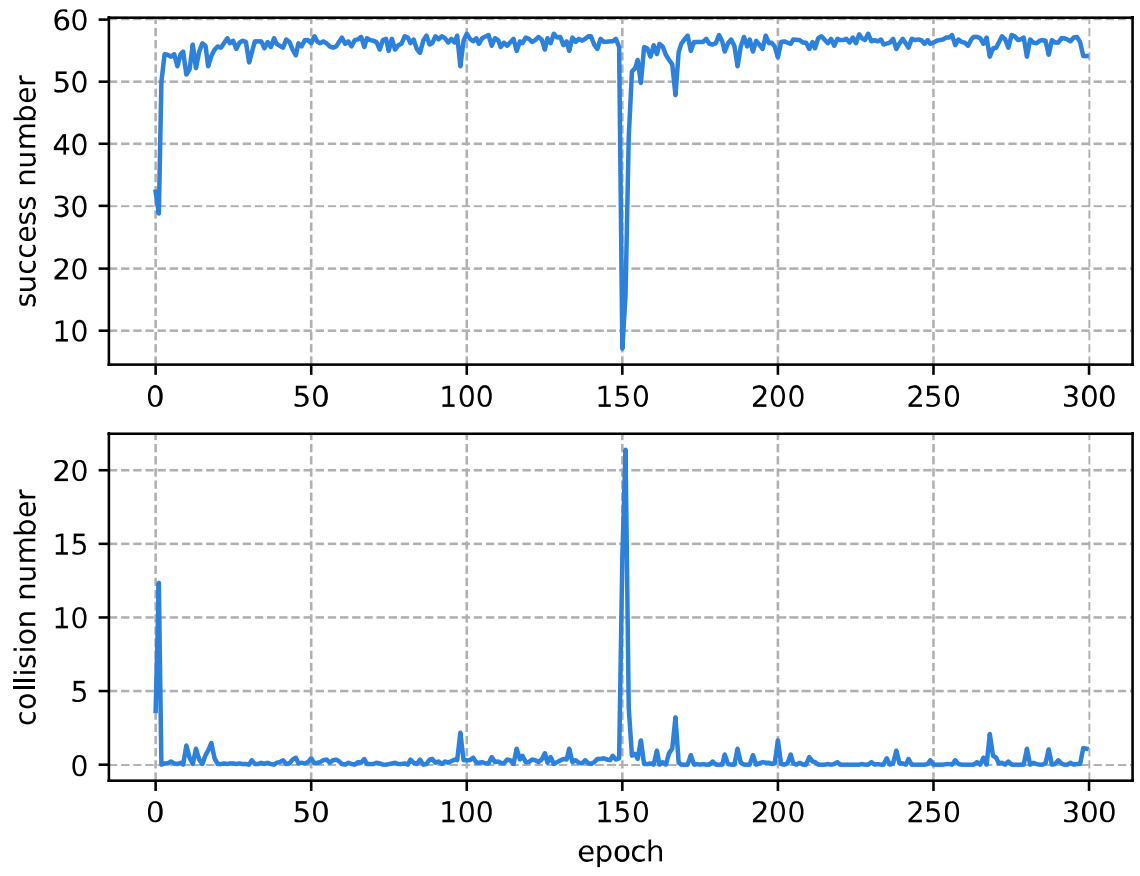

Fig. 14. Three cognitive users with two sensing channels from Markov channels to Markov channels.

### *F. Real-world channels*

In this subsection, we train and evaluate the proposed QMIX-DSA algorithm with the real-world data trace which is downloaded from website [33]. The real-world data trace is produced by an indoor low-power wireless network testbed which is deployed by University of Southern California (USC). It features three generations of wireless sensor nodes, all with IEEE 802.15.4-compliant radios operating on the 2.4 GHz frequency band. The sensor nodes perceive the interference coming from surrounding Wi-Fi networks and record the states of channels.

We implement the QMIX-DSA algorithm for three cognitive users with one sensing channels with the real-world data trace. Meanwhile, we implement the DQSA algorithm in the same wireless setting. In an episode comprised of 20 time slots, three cognitive users with continuous transmission need 60 idle time slots to transmit. From Fig. 15, we observe that the proposed QMIX-DSA algorithm and DQSA converge after certain training steps, but the QMIX-DSA algorithm achieves better performance than DQSA. The experiments result shows that the proposed QMIX-DSA algorithm is effective in real-world wireless setting.

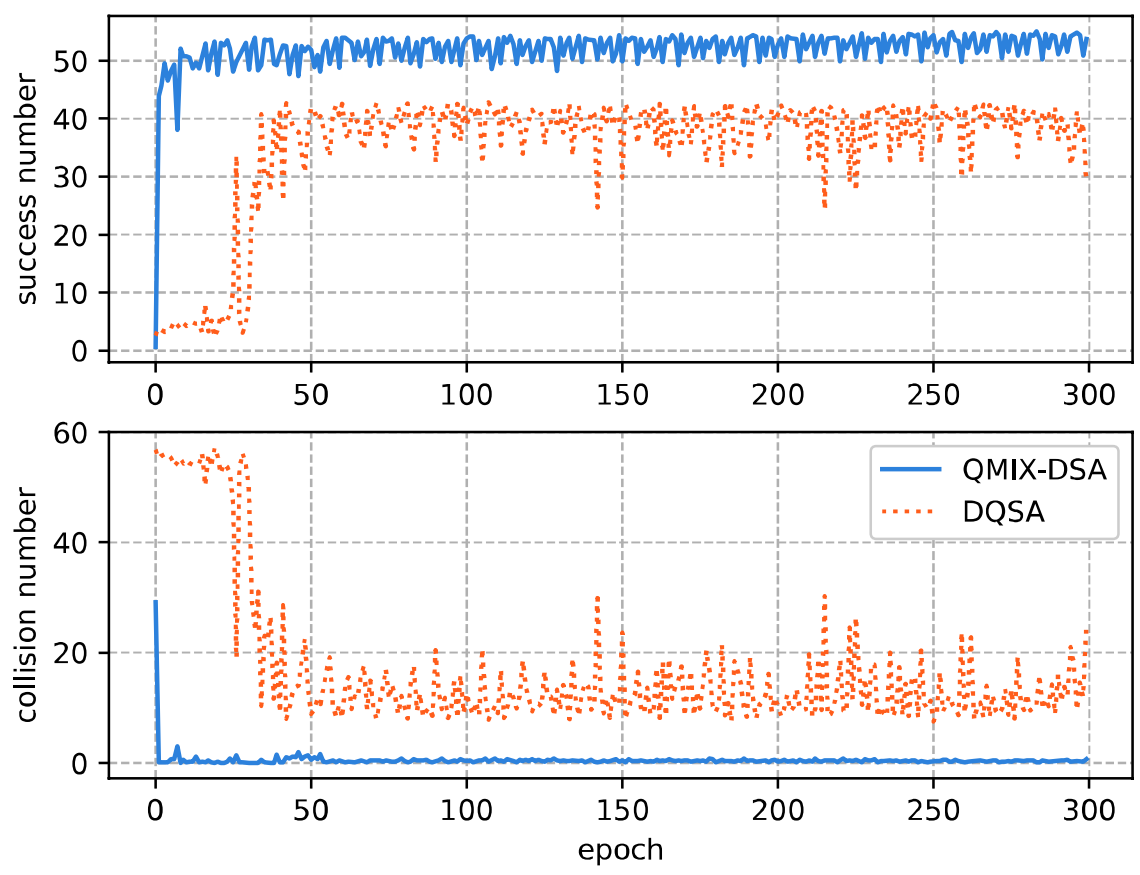

Fig. 15. Three cognitive users with two sensing channels in real-world channels.

## VI. Conclusion

In this paper, we have investigated the distributed DSA problem of multi-user in multi-channel cognitive radio networks. We modeled this problem as a Dec-POMDP, in which the cognitive users have incomplete information of the wireless environment state and the network dynamics is not known to cognitive users prior. The cognitive users learn the network dynamics by interaction with the wireless environment continuously and infer the environment state from their own local observations and historical feedback of the actions. We formulated this problem as a Markov game. Each cognitive user is modeled as an independent learning agent observing its local environment state to make the optimal access strategy autonomously, without exchanging coordination information with the other cognitive users. The goal is to learn a distributed multi-channel access strategy with high channel utilization but low collision rate for multi-user cognitive radio networks.

To overcome the dimension explosion and instability of the multi-agent system, we have proposed a CTDE framework for distributed DSA problem based on MARL. In the training phase, the cognitive users send their observation-action function to the centralized trainer deployed on the edge computing server to learn a cooperative strategy. In the execution phase, the cognitive users make access decisions autonomously according to their local observations. There is no coordination information exchange between the cognitive users, and this reduces the communication overhead greatly. We validated the proposed QMIX-DSA algorithm in various environment settings and compared with the state-of-the-art DQSA algorithm. From the simulation results, we observed that the proposed algorithm outperforms the DQSA algorithm in terms of convergence speed and achieved performance.